\documentclass[twocolumn,superscriptaddress,showpacs,prl,aps,amssymb]{revtex4}
\usepackage{graphicx}
\usepackage{bm}

\newcommand{\Madm}{M_{\rm ADM}}
\newcommand{\MK}{M_{\rm K}}
\newcommand{\beq}{\begin{equation}} 
\newcommand{\eeq}{\end{equation}} 
\newcommand{\beqn}{\begin{eqnarray}} 
\newcommand{\eeqn}{\end{eqnarray}} 
\newcommand{\pa}{\partial}
\newcommand{\na}{\nabla}
\newcommand{\gab}{g_{\alpha\beta}}

\newcommand{\gabd}{g_{\alpha\beta}}

\newcommand{\gmabd}{\gamma_{ab}}
\newcommand{\tgmabu}{\tilde\gamma^{ab}}
\newcommand{\tgmabd}{\tilde\gamma_{ab}}

\newcommand{\fabd}{f_{ab}}

\newcommand{\albe}{\alpha\beta}

\newcommand{\Tba}{T_\alpha{}^\beta}

\newcommand{\zD}{{\raise1.0ex\hbox{${}^{\ \circ}$}}\!\!\!\!\!D}
\newcommand{\alone}{{\raise0.5ex\hbox{${}^{\ 1}$}}\!\!\!\!\alpha}
\newcommand{\Od}{{O}}

\newcommand{\Lie}{\mbox{\pounds}}

\newcommand{\compa}{(M/R)_\infty}

\newcommand{\nalam}{\mathrel{\raise0.9ex\hbox{$^\lambda$}\mkern-14mu
\lower0.0ex\hbox{$\nabla$}}}

\begin{document}  

\title{
Binary neutron stars: Equilibrium models beyond spatial conformal flatness
}  
\author{K\=oji Ury\=u}
\affiliation{
Department of Physics, University of Wisconsin-Milwaukee, P.O. Box 413,  
Milwaukee, WI 53201}
%
%
\author{Fran\c{c}ois Limousin} 
\affiliation{
Laboratoire de l'Univers et de ses Th\'eories, UMR 8102 du C.N.R.S.,
Observatoire de Paris, F-92195 Meudon Cedex, France}
\author{John L. Friedman} 
\affiliation{
Department of Physics, University of Wisconsin-Milwaukee, P.O. Box 413,  
Milwaukee, WI 53201}
\author{Eric Gourgoulhon}
\affiliation{
Laboratoire de l'Univers et de ses Th\'eories, UMR 8102 du C.N.R.S.,
Observatoire de Paris, F-92195 Meudon Cedex, France}
\author{Masaru Shibata}
\affiliation{
Department of Earth Science and Astronomy, Graduate School of Arts and  
Sciences, University of Tokyo, Komaba,  
Meguro, Tokyo 153-8902, Japan}
\date{\today}

\begin{abstract} 
Equilibria of binary neutron stars in close circular orbits
are computed numerically in a waveless formulation: 
The full Einstein-relativistic-Euler system is solved on an 
initial hypersurface to obtain an asymptotically flat form of 
the 4-metric and an extrinsic curvature whose time derivative 
vanishes in a comoving frame. Two independent numerical codes 
are developed, and solution sequences that model inspiraling binary 
neutron stars during the final several orbits are successfully computed.  
The binding energy of the system near its final orbit deviates from 
earlier results of third post-Newtonian and of spatially conformally 
flat calculations.  The new solutions may serve as initial data for 
merger simulations and as members of quasiequilibrium sequences  
to generate gravitational wave templates, and may improve estimates of   
the gravitational-wave cutoff frequency set by the last inspiral orbit.
\end{abstract} 

\maketitle
 
\paragraph{Introduction:}
\label{intro} 

Equilibria of close binary neutron 
stars in circular orbits, constructed numerically, 
have been studied as a model of the final several 
orbits of binary inspiral prior to merger 
(see \cite{BaumgS03} for a review).  
These numerical solutions have been used as 
 initial data sets for merger simulations
\cite{SUTall};
in quasi-equilibrium sequences, to estimate gravitational waveforms 
\cite{SU01,DU02}; and to determine  
the cutoff frequency of the inspiral waves  
\cite{IWMsequence}.

To maintain equilibrium circular orbits in general relativity
one must introduce an approximation that eliminates 
the back reaction of gravitational radiation. 
An ansatz of this kind is the waveless approximation 
proposed by Isenberg \cite{ISEN78}.  
One of his proposals was to choose a conformally flat spatial 
geometry maximally embedded in a spacetime.  
As a result the gravitational field is no longer
dynamical; field equations for the metric components become
elliptic equations.
Wilson and Mathews later rediscovered this waveless 
approximation and applied it to numerical computations of 
binary inspirals \cite{WMM959}.  

Although the Isenberg-Wilson-Mathews (IWM) formulation 
has been widely used for modeling binary 
neutron star and binary black hole inspiral in the past decade 
\cite{WMM959,BA97,BGM989,USE00,GGB02_CookP04,IWMsequence,FGR04}, 
the error associated with its conformally flat 
3-geometry was studied only for stationary axisymmetric systems 
\cite{WM_TEST}.
In models of binary neutron stars, the estimated 
error in the orbital angular velocity, $\Omega$, 
is several percent \cite{ASF96,SU01}, 
implying a comparable deviation from circular orbits 
\cite{BNS_TEST}.
New waveless formulations, incorporating a generic form of the metric,
are suitable for accurate computation of binary 
compact objects \cite{SG04,SUF04}.  
In this letter, we present the first results of numerical 
computations for binary neutron stars modeled in 
one of these formulations \cite{SUF04}.

\paragraph{Formulation of the waveless spacetime:}
\label{sec:II}

The new formulation \cite{SUF04} exactly solves the Einstein-Euler 
system written in 3+1 form on a spacelike hypersurface.  
We follow notation \footnote{Indices $a-d$ and $\alpha-\delta$ are 
abstract, $i,j$ concrete} used in \cite{SUF04}.
The spacetime ${\cal M}={\mathbb R}\times \Sigma$ is foliated by 
the family of spacelike hypersurfaces, $\Sigma_t=\{t\}\times\Sigma$. 
The future-pointing normal $n^\alpha$ to $\Sigma_t$ 
is related to the timelike vector $t^\alpha$ 
(the tangent ${\bm \partial}_t$ to curves $t\rightarrow (t, x), x\in\Sigma$) by 
$t^\alpha = \alpha n^\alpha + \beta^\alpha$, where 
$\alpha$ is the lapse, and where the shift $\beta^\alpha$ 
satisfies $\beta^\alpha n_\alpha=0$.  
A spatial metric $\gmabd(t)$ defined on $\Sigma_t$ 
is equal to the projection tensor 
$\gamma_{\albe} = \gab+n_\alpha n_\beta$ restricted to $\Sigma_t$.  
In terms of a conformal factor $\psi$ and a conformally rescaled spatial 
metric $\tgmabd = \psi^{-4} \gmabd$, 
the metric $\gabd$ takes the form, 
$
ds^2=-\alpha^2dt^2+\psi^4\tilde \gamma_{ij}(dx^i+\beta^i dt)(dx^j+\beta^j dt), 
$
in a chart $\{t,x^i\}$. 
A condition to specify the conformal 
decomposition is $\det\tgmabd=\det\fabd$, where $\fabd$ is 
a flat metric. 

In our waveless formulation we impose, as coordinate conditions, 
maximal slicing  ($K=0$) and the spatially transverse condition 
$\zD_b\tgmabu=0$ (the Dirac gauge \cite{SUF04,BGGN04}), 
where $\zD_b$ is the covariant derivative 
with respect to the flat metric. We then 
restrict time-derivative terms in this gauge 
to guarantee that all components of the field 
equation are elliptic equations, and hence that  
all metric components, including the spatial metric,
have Coulomb-type fall off \cite{SUF04}. While
it is found to be sufficient to impose the condition 
$\pa_t\tgmabu=\Od(r^{-3})$ to have Coulomb-type fall off 
in the asymptotics, 
we impose a stronger condition: $\pa_t\tgmabu=0$.  
For the other quantities, we impose helical symmetry: 
spacetime and fluid variables are dragged 
along by the helical vector $k^\alpha = t^\alpha + \Omega \phi^\alpha$.  
For example, the time derivative of extrinsic 
curvature $K_{ab}$ is expressed as
$\pa_t K_{ab} = -\Lie_{\Omega{\bm\phi}}K_{ab}$.  
The resulting field equations are solved on a slice $\Sigma_0$.  
The Hamiltonian constraint, momentum constraint, spatial trace and 
spatial tracefree part of the Einstein equation are, respectively, 
regarded as elliptic equations for $\psi$, $\beta^a$, 
$\alpha$ and $h_{ab}:=\tgmabd - \fabd$, while 
the extrinsic curvature, $K_{ab}$, 
for this formulation 
is computed from the metric components, 
$K_{ab}=\frac1{2\alpha}\Lie_\beta\gmabd +\frac1{3\alpha}\gmabd D_c(\Omega\phi^c)$.

To compute the motion of binary neutron stars in circular orbits, 
the flow field is assumed to be stationary in the rotating frame.  
Since any solution to the waveless formulation satisfies 
all constraint equations, it is, in particular, an initial data set 
for the Einstein-Euler system.  When one evolves such 
a binary neutron star solution by integrating the Einstein-Euler system, 
the orbits will deviate from exact circularity because of the radiation 
reaction force.  Instead, one can construct an artificial spacetime with 
circular orbits by dragging the waveless solution on $\Sigma_0$ along 
the vector $k^\alpha = t^\alpha + \Omega \phi^\alpha$, 
so that the spacetime has helical symmetry.  Although the spacetime 
so constructed will not exactly satisfy Einstein's equation, 
a family of such spacetimes, associated with circular orbits 
of decreasing separation, will model the inspiral of a binary neutron 
star system during its final several orbits.  
Explicit forms of all equations for the fields and the matter 
are found in \cite{BGGN04,SUF04}.
\paragraph{Numerical methods:}

We have developed two independent numerical schemes to compute 
binary neutron star solutions.  One is based on a finite difference 
method \cite{USE00}, the other one on a spectral method implemented
via the C++ library {\sc Lorene} \cite{Lorene}. 
Detailed convergence tests and 
calibration of each method will be published separately.  
In this letter, we show quantitative agreement 
of the two methods for $h_{ab}$, which is the significant and 
reliable calibration for the new numerical solutions.

In both methods, equations are written in Cartesian coordinate components, 
and they are solved numerically on spherical coordinate grids, 
$r$, $\theta$, and $\phi$.  
In the finite difference method, an equally spaced grid is used from the 
center of orbital motion to $5R_0$ where there are $n_r=16$, $24$, and 
$32$ grid points per $R_0$; from $5R_0$ to $10^4R_0$ a 
logarithmically spaced grid has $60, 90$, and $120$ points 
(depending on the resolution). Here $R_0$ is 
the geometric radius of a neutron star along a line passing through 
the center of orbit to the center of a star.  
Accordingly, for $\theta$ and $\phi$ there are $32$, $48$, and $64$ 
grid points each from $0$ to $\pi/2$ \cite{USE00}.   
For the spectral method, eight domains (a nucleus, six shells and a 
compactified domain extending up to infinity) around each star are used.
In each domain, the number of collocation points is chosen to be 
$N_r\times N_{\theta} \times N_{\phi} = 25 \times 17 \times 16$ 
and $33 \times 21 \times 20$ \cite{BGM989}.

%
\begin{figure}
\begin{tabular}{cc}
\begin{minipage}{.5\hsize}
\begin{center}
\includegraphics[height=40mm,clip]{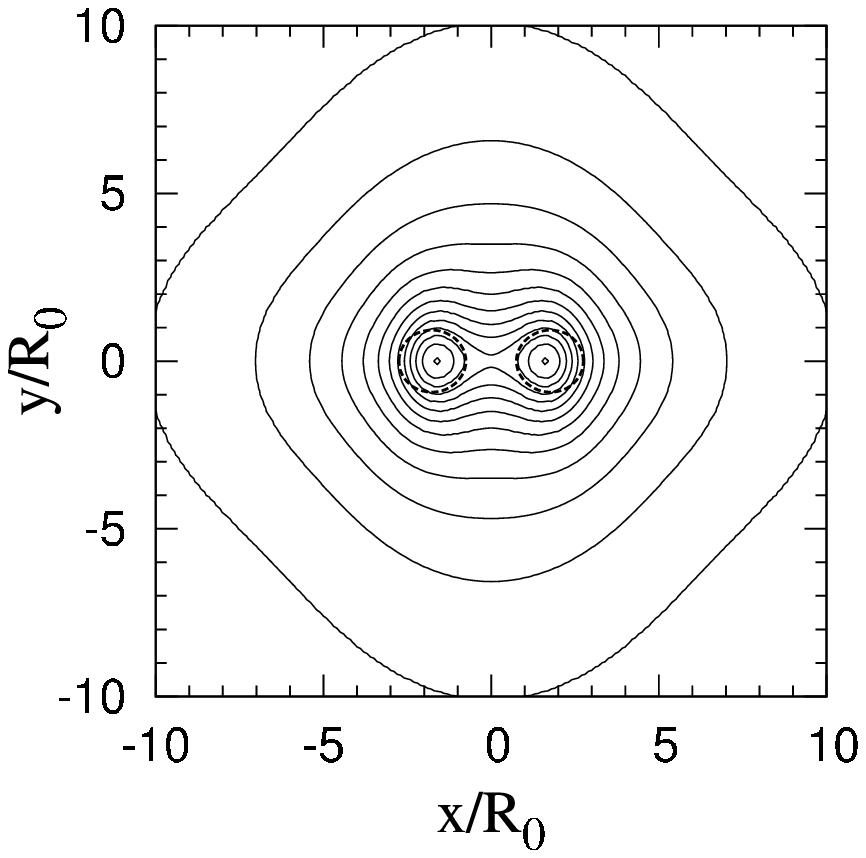}
\end{center}
\end{minipage} 
&
\begin{minipage}{.5\hsize}
\begin{center}
\includegraphics[height=37mm,angle=-90,clip]{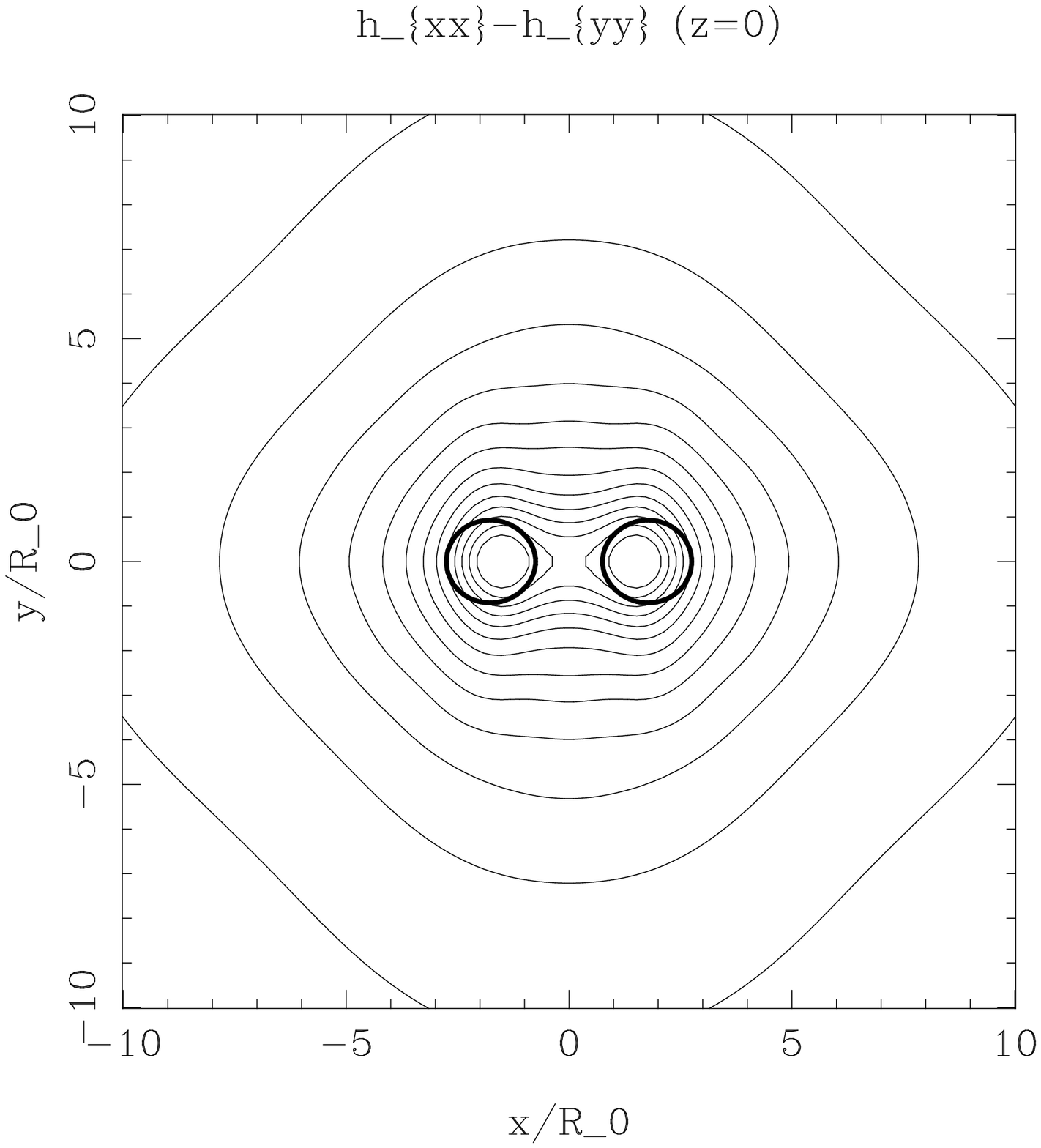}
\end{center}
\end{minipage} 
\end{tabular} 
\caption{Contours of 
$(h_{xx}-h_{yy})/2$ in the $xy$-plane, computed by 
the finite difference code (left) and 
by the spectral code (right). 
The binary separation $a$ is given by $a/R_0 = 3.5$.  
Contours extend from $-0.014$ to $-0.002$ with step $0.001$. 
}
\label{fig:hab_contour}
\end{figure}
%
%
%
\begin{figure}
\begin{center}
\includegraphics[scale=0.28,clip]{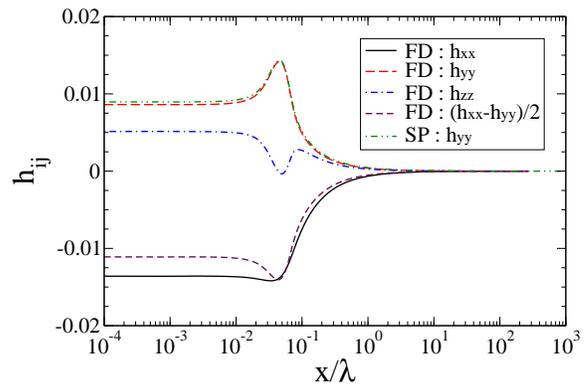}\\[2mm]
\caption{Components $h_{ij}$ along the $x$-axis, 
normalized by $\lambda=\pi/\Omega$.  
A neutron star extends from $x/\lambda = 0.02024$ to $0.07422$.  
Curves labeled FD and SP display results of the finite difference 
and spectral codes, respectively.  
}
\label{fig:xaxis}
\end{center}
\end{figure}

\paragraph{Numerical solutions for binary neutron stars:}

A model of the evolutionary path of binary inspiral is 
given by a sequence of equilibria  
along which the neutron star matter is assumed 
to be isentropic; and the implied fluid flow is assumed to 
conserve the baryon number, entropy and 
vorticity of each fluid element \cite{KBC92,BaumgS03}.  
In the case where the spins of component stars 
are negligible, the flow becomes irrotational; one can introduce 
the velocity potential $\Phi$ by $hu_\alpha = \na_\alpha\Phi$, 
where $h$ is the specific enthalpy and 
$u^\alpha$ is the fluid 4-velocity.  
For isentropic flow, one can assume a one-parameter equation 
of state, $p=p(\rho)$, with $\rho$ the baryon mass density.
The matter is then described by two independent variables, a thermodynamic variable 
such as $p/\rho$, and the velocity potential $\Phi$.  
In this letter, we assume a polytropic equation of state 
$p=\kappa \rho^\Gamma$ with adiabatic index $\Gamma=2$, 
and we display results for equal-mass binaries with 
the rest mass of each star to be that of 
a single spherical star of compactness $\compa = 0.17$.
(Note: The maximum compactness 
of a spherical star for this equation of state is $\compa = 0.216$.
The compactness $\compa$ is defined as the ratio of graviational mass 
to circumferential radius of an isolated spherical star with 
the same rest mass.)

In Fig.~\ref{fig:hab_contour}, contours
of the components $h_{ij}$ computed by the two numerical 
codes are shown for selected solutions.  
In these solutions, the separation in coordinate distance 
between the coordinate center of each neutron star 
is set to $a/R_0 = 3.5$.  
From these contours, one can verify qualitative agreement of 
the results from the two independent numerical methods.  
In Fig.~\ref{fig:xaxis}, components $h_{ij}$ along the $x$-axis 
are plotted for the same solution, where the $x$-axis passes through 
the centers of the neutron stars.  
Precision of integral quantities characterizing the solutions 
is shown by the finite difference (spectral) method comparisons: 
$\Omega M_\infty = 0.03565$ $(0.03565)$, 
$\Madm/M_\infty = 0.98825$ $(0.98826)$, and 
$J/M^2_\infty = 0.9212$ $(0.9165)$. 
%

\begin{figure}
\begin{center}
\includegraphics[scale=0.28,clip]{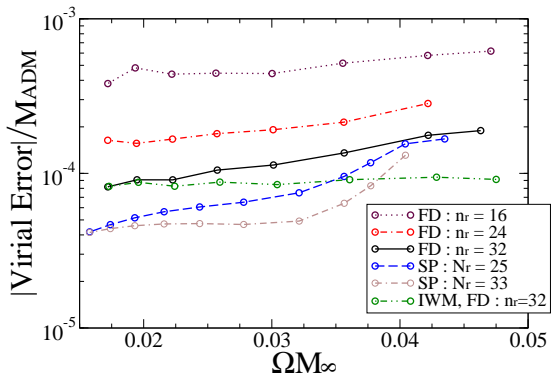}
\caption{Virial error vs. angular velocity $\Omega$, normalized by $M_\infty$, 
twice the gravitational mass of an isolated neutron star.  
Each curve labeled FD shows results of a finite difference 
code with a given resolution.  Curves labeled SP and IWM show 
results of the spectral code and the spatially conformally flat approximation, 
respectively.}
\label{fig:virial}
\end{center}
\end{figure}

In \cite{SUF04}, it is shown that 
the ADM mass, $\Madm$, and the asymptotic Komar mass, 
$\MK$ are equal, $\Madm=\MK$, under asymptotic conditions satisfied 
by the solutions in the present formulation.  
The equality is related to a virial 
relation for the equilibrium, 
\beq
\int x^i \gamma_i\!{}^\alpha \nabla_\beta\Tba \sqrt{-g}d^3x =0, 
\label{eq:virial}
\eeq
that we use 
to evaluate the accuracy of numerical solutions.  
Fig.~\ref{fig:virial} shows the computed value of the  
virial integral in Eq.~(\ref{eq:virial}), normalized by $\Madm$,
along the sequence.
We also evaluated $\Madm$ and $\MK$ 
each defined by the surface integral in the asymptotics, 
and confirmed that, 
for each model, the difference of the two masses is no larger than 
$|\Madm-\MK|/\Madm$ $ \sim 0.01 \%$ for the finite difference method 
and $\sim 0.001 \%$ for the spectral method; these errors are consistent with 
the numerical errors of the virial relation shown in Fig.~\ref{fig:virial}.

\begin{figure}
\begin{center}
\includegraphics[scale=0.28,clip]{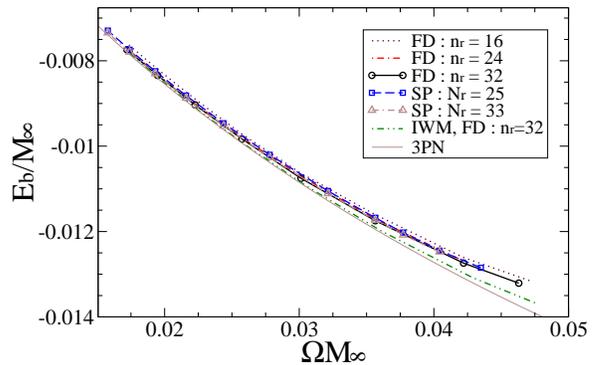}
\caption{Binding energy $E_b:=\Madm-M_\infty$ 
with respect to the normalized angular velocity along
the sequence.
Curves are labeled as in Fig.~\ref{fig:virial}.  
The thin solid curve corresponds to 
the third order post-Newtonian 
calculation \cite{Blanc02}.}
\label{fig:bindE-seq}
\end{center}
\end{figure}
\begin{figure}
\begin{center}
\includegraphics[scale=0.65,clip]{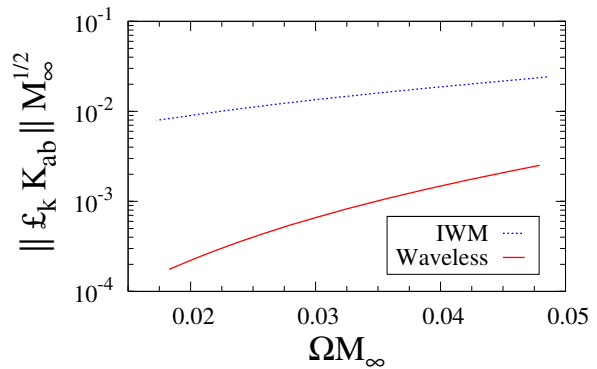}
\caption{Norm of Lie derivative of the extrinsic curvature along
the helical vector [cf. Eq.~(\ref{eq:norm})] 
with respect to the normalized angular velocity. 
}
\label{fig:norm-seq}
\end{center}
\end{figure}


%
In Fig.~\ref{fig:bindE-seq}, 
the binding energy $E_b = \Madm - M_\infty$ 
along the sequence is plotted and compared with that
resulting from a third 
post-Newtonian (3PN) calculation \cite{Blanc02}
and  IWM formulation.
The waveless sequence fits the 3PN curve 
well at larger separation, 
and reaches a configuration with 
a cusp without any turning point in the binding energy curve,
in agreement with results of the IWM formulation \cite{USE00,BGM989}  
(the spectral code does not yet converge 
for the closest orbits -- largest $\Omega M_\infty$-- 
of Figs.~\ref{fig:virial} 
and \ref{fig:bindE-seq}, because it is more sensitive to tidal deformation: 
higher multipoles in the density of each star lead to a divergent iteration). 
The binding energy $E_b$ of the waveless sequences clearly deviates 
from that of the 3PN and IWM sequences at the larger values of 
$\Omega M_\infty$.  
This suggests that the 3PN and IWM formulations each overestimate 
the binding energy -- in the 3PN case, by neglecting the tidal 
deformation, in the IWM formulation by neglecting the contribution
from $h_{ab}$.  

Finally, to estimate the deviation of the orbit from circularity, 
we evaluate the formal expression for the extrinsic 
curvature of a solution with exact helical symmetry, 
(the case for which the time-evolved data has an exactly circular orbit), 
$\hat K_{ab} = \frac1{2\alpha}\Lie_{\beta+\Omega\phi}\gmabd$.
Because $\Lie_k \hat K_{ab}$ vanishes for exact helical symmetry, its 
norm, defined on the support $V$ of the fluid, 
\beq
||\, \Lie_k \hat K_{ab}\, || 
:=\left[\int_V \gamma^{ac}\,\gamma^{bd}\, 
\Lie_k \hat K_{ab}\,\Lie_k \hat K_{cd} \sqrt{\gamma}\,d^3x\right]^{1/2} \!\!\!\!,
\label{eq:norm}
\eeq
is a measure of the deviation from circularity.

Fig.~\ref{fig:norm-seq} shows that, for all separations,
the values of $||\, \Lie_k \hat K_{ab}\, ||$
for the waveless solutions are more than 
an order of magnitude smaller than those of IWM solutions.  
The result supports the expectation that IWM data enforces 
circularity with significantly less accuracy than the 
corresponding waveless solutions, even for larger separation.
This may be important: Even for a sudden turn-on of
radiation-reaction a post-Newtonian analysis \cite{BNS_TEST}
shows eccentricity $<1.5\%$ at $\Omega M < 0.03$ for initially circular
orbits, and the gravitational radiation reaction should be more gradual 
for our waveless data sets.

\paragraph{Discussion:}
\label{sec:IV}

In 2nd post-Newtonian theory (e.g.~\cite{ASF96}), the
correction $\Delta E_b$ to the binding energy due to the contribution
of $h_{ab}$ is of order $M_{\infty}h_{ab}v^a v^b$, where the magnitude 
of orbital velocity, $v^a$, may be typically $v \approx
0.34(\Omega M_{\infty}/0.04)^{1/3}$. Since $h_{ab}$ is $O(v^4)$, 
$\Delta E_b/M_{\infty} = O(v^6) \sim 10^{-3}$ for
$\Omega M_{\infty} \sim 0.04$. This agrees with the 
difference between the binding energies calculated by the IWM and
waveless formulations in Fig.~\ref{fig:bindE-seq}.

The quantity $dE_b/d \Omega$ is important for the data analysis of 
gravitational waves, because it determines the evolution of the 
wave's phase, $\Phi_{\rm GW}=2\int \Omega(t) dt$.  
In adiabatic evolution, the time dependence of 
angular velocity $\Omega(t)$ is calculated from 
$d\Omega/dt=|(dE/dt)_{\rm GW}|/(dE_b/d\Omega)$, 
where $(dE/dt)_{\rm GW}$ is the luminosity of gravitational waves.  
Our present result shows that the derivative $dE_b/d\Omega$ 
of waveless sequences is $\sim 10$--15\% larger than 
those of IWM and 3PN curves for $\Omega M_{\infty}\agt 0.035$.  
Since $\sim 2$ orbits are maintained from $\Omega M_{\infty} = 0.035$ 
to merger for the case with $(M/R)_{\infty}=0.17$ \cite{SU01}, 
the error in the IWM and 3PN values of $\Phi_{\rm GW}$ would 
accumulate to $\sim 50\%$ over the last $\sim 2$ orbits. 
The phase error leads to error during the final orbits before merger
of the computed frequency, whose final behavior constrains the 
equation of state of nuclear matter \cite{ZCM94,IWMsequence}. 
Waveless solutions may determine phase and frequency with significantly 
greater accuracy -- particularly if, to overcome  
radial-motion error, one first calibrates the frequencies of a set of 
quasiequililbrium sequences, using (for example) time evolutions.

Phase error may be much larger for the final orbits 
of binary black hole and black hole--neutron star inspirals.  
In these cases, $\Omega M_{\infty}$ in the last orbit may reach 
or exceed $0.1$ (e.g.~\cite{GGB02_CookP04,Blanc02}).  
Since $\Delta E_b$ is of order $O(v^6)$,
the phase error is likely to be of order unity for 
$\Omega M_\infty \agt 0.1$.  Therefore, a template constructed 
from the IWM formulation may cause a systematic error 
in the data analysis.  
Our waveless approximation may improve binary black hole
and black hole--neutron star solutions for this purpose.

\paragraph{Acknowledgement:}
This work was supported by 
Monbukagakusho Grants No.~17030004 and 17540232, 
and NSF grants No.~PHY0071044 and PHY0503366.
KU thanks l'Observatoire de Paris for 
financial support and 
N. Kanda at Osaka City University.

\end{document}